**Polynomial-time-scaling quantum dynamics with time-dependent quantum Monte Carlo**

Ivan P. Christov

Physics Department, Sofia University, 1164 Sofia, Bulgaria

**Abstract**

Here we study the dynamics of many-body quantum systems using time dependent quantum Monte Carlo method where the evolution is described by ensembles of particles and guide waves. The exponential-time scaling inherent to the quantum many-body problem is reduced to polynomial-time computation by solving concurrently a set of coupled Schrödinger equations for the guide waves in physical space and a set first order equations for the Monte Carlo walkers. We use effective potentials to accounts for the local and nonlocal quantum correlations in time-varying fields, where for fermionic states an exchange "hole" is introduced explicitly through screened Coulomb potentials. The walker distributions for the ground states of para- and ortho-helium reproduce well the statistical properties, such as the electron-pair density function, of the real atoms. Our predictions for the dipole response and the ionization of an atom exposed to strong ultrashort optical pulse are in good agreement with the exact results.

**Introduction**

The advent of accurate non-perturbative time-dependent methods is indispensable to meet the challenge brought by the rapid progress in femtosecond and attosecond experimental techniques. Although the correlated electron motion in atoms and molecules is correctly described by the many-body time-dependent Schrödinger equation (TDSE), its direct numerical solution remains prohibitively time-expensive even for parallel classical computer because the computational cost scales exponentially with system size. For example, the numerical solution of TDSE for N particles on a grid of K nodes in each of the three spatial dimensions would require workload proportional to $M^{3K}$ that concerns both the number of floating point operations and the memory requirements. It is generally believed that exponential-time scaling is necessitated in order to account for the essential non-local quantum correlation effects in a quantum system described by complex wave function $\Psi(\mathbf{r}_1,...,\mathbf{r}_N,t)$ which resides in configuration space. Nevertheless, several techniques have been developed to lessen the workload by reducing the many-body TDSE to a set of lower-dimension equations, which can treat time-dependent quantum problems approximately, with a time-polynomial cost. The most widely used among those are the self-consistent time-dependent methods, encompassing time-dependent Hartree-Fock (TDHF) [1,2] and time-dependent Density Functional theory (TDDFT) [3,4], where the former ignores the dynamic Coulomb correlations between the electrons while the latter reduces the many-body problem to single-body problems of non-interacting electrons moving in an effective exchange-correlation potential. The Dirac–Slater formula is commonly used to estimate the exchange part of the interaction in TDDFT [5,6], while closed formulas to approximate the correlation interaction of the electrons are introduced [7]. However, since optimized for the ground state of the system, the exchange-correlation potential is not generally believed to be reliable for time dependent processes where significant deformation of the electron distribution may take place. Also, due to the semi-empirical character of the exchange-correlation potential, TDDFT suffers hidden self-interaction burden which can modify uncontrollably the outcome from the quantum calculations. Other wave-function based methods invoke series expansions over multiple configurations (Slater determinants) in order to account for the electron exchange and correlation effects in a TDHF manner [8]. Although still time-polynomial the efficiency of these methods is hampered by the calculation of large number of Coulomb and exchange integrals.

It is well established that the modulus square of the quantum wevefunction represents the probability distribution for an *infinite* ensemble of replicas (realizations) of the electron in position space. In this sense even TDHF and TDDFT cannot be considered to be true single-body theories. However, if we preserve the probabilistic interpretation of quantum mechanics but assume a *finite* (countable) number of classical particles to represent the electron probability distribution in physical space, we may reduce the calculation workload from exponential to polynomial while retaining the fully correlated many-body quantum dynamics. To this end, we need a reliable mathematical algorithm for dynamically mapping discrete distributions to continuous ones, which would compensate for the lack of knowledge of the exact positions of all replicas described by the wavefunctions while using finite number of classical particles. This strategy implies a Monte Carlo (MC) type of calculation for the time evolution of classical particles (walkers) in conjunction with the evolution of quantum waves in physical space, where each individual MC walker samples its own distribution given by the modulus square of the corresponding 3D guide wave. Unfortunately, the known quantum Monte Carlo (QMC) techniques [9] do not allow treatment of time-dependent processes, that is in part because QMC uses branching which is an essentially random process that would violate the causality of the Schrödinger equation for any finite ensemble of walkers which propagate in real time. On the other hand, relying on stochastic models alone, such as those suggested by Nelson [10] or Bohm-Vigier [11], it is not possible to recover all features of quantum theory [12]. Recently a new, time-dependent quantum Monte-Carlo (TDQMC), methodology was proposed that combines self-consistently the motion of a set of Monte Carlo walkers with the evolution of ensemble of quantum waves in physical space [13]. Although guided by the first-order de Broglie-Bohm equation [14,15], the walkers used in this method do not represent individual electrons. Instead, an assumption is adopted that the walker distribution in space corresponds to the electron density function which, in particular, allows the walkers to be at rest for certain stationary states, without violating the uncertainty principle. At the same time, the infamous "fermion sign" problem is avoided in TDQMC because the electron density function, which is sampled by the walkers, is positively defined everywhere. The guide waves in this model evolve according to a set of coupled time-dependent Schrödinger equations in physical space. These waves retain their statistical interpretation, which is in contrast to in the de Broglie theory where the guiding waves are ascribed to physical fields. In this way the TDQMC method reconciles some aspects of the standard (Copenhagen) and the de Broglie-Bohm theories, with direct application to quantum calculations where explicitly correlated wave functions converge faster than using multiple configurations (see e.g. [16]).

Schematically, the TDQMC algorithm involves a number of steps[13] that include initialization of separate Monte Carlo ensembles of walkers and corresponding guide waves, followed by their concurrent propagation in complex time toward steady state, in presence of random component in walker's motion to simulate the quantum drift and diffusion. The time dependent quantum dynamics is studied by switching to real time after the ground state of the system is established, where interaction with external electromagnetic fields and/or with other particles can be encompassed. This formulation ascertains that the walker density corresponds to the electron density in space at all times. In this paper we discuss the possibility to formulate the TDQMC method by using reduced number of Slater determinants for fermionic states. It is demonstrated that this approach further improves the efficiency of the calculations of correlated quantum dynamics.

**Screened potential formulation of TDQMC**

Here we consider correlated electron dynamics in multi-electron atoms subjected to an external electromagnetic field, although the same method can be applied to molecules and more complex objects [13]. For fixed-nuclei, the system of $N$ electrons is described by the many-body Schrödinger equation:

$$i\hbar \frac{\partial}{\partial t}\Psi(\mathbf{R},t) = -\frac{\hbar^2}{2m}\nabla^2 \Psi(\mathbf{R},t) + V(\mathbf{R})\Psi(\mathbf{R},t) , \qquad (1)$$

where $\mathbf{R} = (\mathbf{r}_1,...,\mathbf{r}_N)$ is a 3N dimensional vector in configuration space which specifies the coordinates of N electrons, and $\nabla = (\nabla_1, \nabla_2,..., \nabla_N)$. The potential term in Eq. (1) is a sum of electron-nuclear, electron-electron, and external potentials:

$$V(\mathbf{r}_1,...,\mathbf{r}_N) = V_{e-n}(\mathbf{r}_1,...,\mathbf{r}_N) + V_{e-e}(\mathbf{r}_1,...,\mathbf{r}_N) + V_{ext}(\mathbf{r}_1,...,\mathbf{r}_N,t)$$

$$= \sum_{k=1}^{N} V_{e-n}(\mathbf{r}_k) + \sum_{k>l}^{N} V_{e-e}(\mathbf{r}_k - \mathbf{r}_l) + V_{ext}(\mathbf{r}_1,...,\mathbf{r}_N,t) . \qquad (2)$$

The TDQMC approach to many-body dynamics assigns M classical walkers to each electron where each walker follows a definite trajectory through de Broglie-Bohm guiding equation for the velocity of the k-th walker form i-th electron ensemble [17, 18]:

$$\mathbf{v}(\mathbf{r}_i^k) = \frac{\hbar}{m} \text{Im} \left[ \frac{1}{\Psi^k(\mathbf{r}_1,...,\mathbf{r}_N,t)} \nabla_i \Psi^k(\mathbf{r}_1,...,\mathbf{r}_N,t) \right]_{\mathbf{r}_j = \mathbf{r}_j^k(t)}, \quad (3)$$

where i=1,...,N; k=1,...,M, and for no spin variables in Schrödinger equation the many-electron wavefunction depends on the individual time-dependent guide waves $\varphi_i^k(\mathbf{r}_i,t)$ through anti-symmetrized products (Slater determinants or sums of Slater determinants), for each configuration:

$$\Psi^k(\mathbf{r}_1,\mathbf{r}_2,...,\mathbf{r}_N,t) = A \prod_{i=1}^{N} \varphi_i^k(\mathbf{r}_i,t). \quad (4)$$

Concurrently, the guide waves obey a set of coupled TDSE:

$$i\hbar \frac{\partial}{\partial t} \varphi_i^k(\mathbf{r}_i,t) = \left[ -\frac{\hbar^2}{2m} \nabla_i^2 + V_{e-n}(\mathbf{r}_i) + \sum_{j \neq i}^{N} V_{e-e}^{eff}[\mathbf{r}_i - \mathbf{r}_j^k(t)] + V_{ext}(\mathbf{r}_i,t) \right] \varphi_i^k(\mathbf{r}_i,t), \quad (5)$$

where each effective electron-electron potential is expressed as a Monte Carlo sum over the smoothed walker distribution:

$$V_{e-e}^{eff}[\mathbf{r}_i - \mathbf{r}_j^k(t)] = \frac{1}{Z_j^k} \sum_{l=1}^{M} V_{e-e}[\mathbf{r}_i - \mathbf{r}_j^l(t)] K\left( \frac{|\mathbf{r}_j^l(t) - \mathbf{r}_j^k(t)|}{\sigma_j^k(\mathbf{r}_j^k,t)} \right), \quad (6)$$

where:

$$Z_j^k = \sum_{l=1}^{M} K\left(\frac{\left|\mathbf{r}_j^l(t) - \mathbf{r}_j^k(t)\right|}{\sigma_j^k\left(\mathbf{r}_j^k,t\right)}\right), \qquad (7)$$

where $K$ is a smoothing kernel and $Z_j^k$ is the weighting factor [13]. The width $\sigma_j^k\left(\mathbf{r}_j^k,t\right)$ of the kernel in Eq.(6) is a measure for the length of nonlocal quantum correlations within the ensemble of walkers that represent the j-th electron. In this way the quantum nonlocality is manifested as a possibility for Coulomb interaction of the k-th walker from the j-th electron ensemble not only with the k-th walkers from the ensembles that represent the rest of the electrons, but also with other walkers from these ensembles that lie within the range of the nonlocal correlation length $\sigma_j^k\left(\mathbf{r}_j^k,t\right)$. In fact, the smoothing kernels in Eq. (6) transform the Coulomb potential due to discrete walker distribution to an effective potential of a continuous charge distribution (infinite number of walkers), where the length $\sigma_j^k\left(\mathbf{r}_j^k,t\right)$ can be determined by performing kernel density estimation [19, 20] over the Monte Carlo data.

An important aspect of the many-body quantum dynamics for fermions is the computation of trajectories through the guiding equation (3). Our calculations reveal that using Slater determinants alone may not be optimal because according to Eq.(4) the function $\Psi^k(\mathbf{r}_1,...,\mathbf{r}_N,t)$ resides in configuration space, while at the same time the Schrödinger equations for the guide waves $\varphi_i^k(\mathbf{r}_i,t)$ (Eq.(5)) operate in physical space. Also, the calculation of a large number of Slater determinants would be needed for all configurations that rest within the nonlocal correlation length. As a result, for N electrons and M walkers for each electron, the calculation time may grow faster than $(NM)^3$. In order to address that challenge we adopt here a formulation based on the assumption that two equal-spin electrons interact through Coulomb potential which is modified due to the exchange effects. It is assumed that the exchange interaction causes fluctuations in the position of the MC walkers such that each walker feels a Coulomb field due to a smeared electron charge centered at the position of the walkers that belong to the rest of the electrons. The role of the smeared electron charge is to prevent the rest of electrons from approaching too much any given same-spin electron. Assuming Gaussian charge distribution of width $r_s$ we find from the solution of Poisson equation that the screened electron-electron repulsion potential is $\sim erf(r/r_s)/r$ where $r_s$ reflects the characteristic size of the "hole" around the electron that arises due to the exchange effects. Notice that an approach to the effects of fermion exchange on the electron repulsion that uses screened Coulomb potentials was proposed by Overhauser [21], although that model cannot be used here directly because the corresponding potential vanishes outside the "hole". Coulomb potentials screened by error function have also been used by other authors for different purposes, e.g. [22, 23], and have also been related to the Ewald summation method which is aimed to reduce the calculation of Coulomb interactions in many-electron systems to $\sim O(N)$ work [24]. Although, formally, the modification of the TDQMC algorithm with using screened electron repulsion potentials is small, it leads to significant improvement and simplification of the calculations. For example, in a single-determinant approximation, the expression for the many-body wave function in Eq.(4) is replaced by a simple product:

$$\Psi^k(\mathbf{r}_1,\mathbf{r}_2,...,\mathbf{r}_N,t) = \prod_{i=1}^{N} \varphi_i^k(\mathbf{r}_i,t), \qquad (8)$$

where the guiding equations for the Monte Carlo walkers simplify to:

$$\mathbf{v}(\mathbf{r}_i^k) = \frac{\hbar}{m} \text{Im} \left[ \frac{1}{\varphi_i^k(\mathbf{r}_i,t)} \nabla_i \varphi_i^k(\mathbf{r}_i,t) \right]_{\mathbf{r}_i = \mathbf{r}_i^k(t)}, \tag{9}$$

which substitutes Eq.(3). At the same time, the coupled TDSE for the guiding waves remain unchanged, but with the electron-electron Coulomb potential in Eq.(6) multiplied by a screening function:

$$V_{e-e}[\mathbf{r}_i - \mathbf{r}_j^l(t)] \to V_{e-e}[\mathbf{r}_i - \mathbf{r}_j^l(t)] erf\left[\frac{|\mathbf{r}_i - \mathbf{r}_j^l(t)|}{r_i^s \delta_{si,sj}}\right] \tag{10}$$

The Kronecker symbol $\delta_{si,sj}$ in Eq.(10) restricts the effect of the exchange hole to only the same-spin walkers, while the value of hole size (screening parameter) $r_i^s$ can be estimated from the corresponding TDHF solution. Here we assume that for few electron atoms the hole size is the same for all electrons, and we denote it as $r_s$. An additional advantage when using Eq.(10) is that the screened potential is not singular for $\mathbf{r}_i \to \mathbf{r}_j^l(t)$, and thus the fermionic wavefunction does not develop Coulomb cusp. The screened potential approach of Eq.(10) can be used in conjunction with using Slater determinants (Eq.(3)) during the preparation of the ground state of the system, in order to better select walker's distribution near the nodal regions.

**Results and discussion**

In order to examine whether the screened electron-electron repulsion can display correctly the correlated electron dynamics for fermionic states we calculate the ground state and the time-dependent response of an atom subjected to a strong femtosecond laser pulse. We focus on the dipole moment and ionization of one-dimensional helium atom in a spin-singlet ground state (para-helium) and in a triplet state (ortho-helium). The preparation of the ground state is described elsewhere [13], as for ortho-helium the guiding waves that belong to different parallel-spin electrons are orthogonalized using Gram-Schmidt procedure. The electron-nuclear and the screened electron-electron interactions are approximated by the following potentials:

$$V_{e-n}(x_i) = -\frac{2e^2}{\sqrt{a+x_i^2}}; \tag{11}$$

$$V_{e-e}[x_i - x_j^k(t)] = \frac{e^2}{b+|x_i - x_j^k(t)|} erf\left[\frac{|x_i - x_j^k(t)|}{r_s \delta_{si,sj}}\right], \tag{12}$$

where the role of the parameters $a$ and $b$ is to prevent the singularity at the origin in one spatial dimension [25]. We compare the predictions from the TDQMC method with the results from the direct numerical solution of TDSE (Eq.(1)) in 2D configuration space for a two-electron atom in symmetric and anti-symmetric ground state (named hereafter "exact solution").

First we explore the ground state properties for 1D para- and ortho-helium atoms for parameters $a=1$ a.u. (atomic units) and $b=0.5$ a.u. in Eq. (11) and Eq. (12), for 1000 Monte Carlo walkers. We assign a separate guide function $\varphi_i^k(x_i, t=0)$ to each walker with coordinate $x_i(t=0)$ where the initial distributions of the walkers are Gaussians with standard deviation σ=1 a.u. After propagation over 400 complex time steps of size (0.1, 0.1) a.u., and in presence of random component in walker's motion [13], the ensembles of walkers and guide waves converge to steady state. For para-helium the effect of charge screening in Eq. (12) is not present and we obtain ground state energy of -2.15 a.u., while for ortho-helium we use hole size $r_s$=0.67 a.u. which yields for ground state energy -1.85 a.u.. These values are within less than 5% from the exact energies obtained by a direct diagonalization of the 2D atomic Hamiltonian. The stationary walker distributions in configuration space for para- and ortho-helium are depicted in Figures 1(a) and 1(b), respectively. It is seen from Fig.1(b) that the walker distribution occupies half of the configuration space available for the fermionic two electron state, which reflects the simplified guiding equation (9). The calculation of the effective electron-electron interaction in Eq. (6) is performed using adaptive kernel density estimation over the walker distributions using Gaussian kernels. First, a pilot density estimate of the walker distribution for the $j$-th electron $\rho_j^k(x,t)$ is calculated using kernel density estimation with constant bandwidth σ. Next, each $\sigma_j^k$ in Eq.(6) is estimated through the formula $\sigma_j^k(x,t) = \sigma\left(G_j / \rho_j^k(x,t)\right)^\alpha$ where $G_j$ is the geometric mean of the values of $\rho_i^k(x,t)$ for all walkers ($k$=1,…,M), and $0 \leq \alpha \leq 1$. Figure 2(a,b) shows the correlation lengths for the two electrons as a function of the distance from the core, for $\alpha = 0.2$. For para-helium the electron density is highest at the core for both electrons where the minimum of nonlocal correlation length (~0.7 a.u.) is reached. For ortho-helium the inner electron is more localized around the core because of both the lack of other same-spin electron in 1S state and the repulsive Coulomb field due to the outer electron. Therefore the minimum nonlocal correlation length in this case is ~0.5 a.u. for the 1S electron while it is close to 2 a.u. for the 2S electron.

The use of Monte Carlo walkers offers the opportunity to calculate various statistical parameters without explicitly referencing to wavefunctions and their integrals. Besides the electron density which is provided directly by the MC output other important parameter that characterizes the electron statistics is the pair density function (PDF), which can be considered to be a histogram of the distances between the electrons, e.g. [3]. Also, PDF is directly related to the experimentally measurable electronic structure factor. In our model PDF is easily calculated using the distance $\left|\mathbf{r}_i^k(t) - \mathbf{r}_j^k(t)\right|$ between the k-th walkers from i-th and the j-th electron ensembles. Figure 3(a) compares, for para-helium ground state, the smoothed PDF obtained by TDQMC as compared to the Hartree-Fock result. The presence of Coulomb hole in the TDQMC curve (blue line) is clearly evidenced, where the walkers are located with higher probability at a distance ~1.7 a.u. from each other, while the Hartree-Fock result shows no Coulomb correlation in walker's position. For ortho-helium (Fig. 3(b)) both curves have maximum close to 2.a.u. that is due to the exchange repulsion of equal spin electrons, though the maximum of the TDQMC curve is shifted to slightly larger distances because of stronger Coulomb repulsion. These results for the ground state are in good correspondence with other calculations that use highly accurate helium wave functions [23, 26].

The Monte Carlo methodology used here allows one to simplify the calculation of quantum averages by reducing these to sums over the walker distributions. For example, the calculation of the dipole moment of the i-th electron can be calculated using guide waves [13]:

$$d_i(t) \propto \sum_{k=1}^{M} \int \left|\varphi_i^k(\mathbf{r},t)\right|^2 \mathbf{r} d\mathbf{r} , \qquad (12)$$

or by using the walker distribution along the trajectories $\mathbf{r}_i^k(t)$, via the substitution:

$$\left|\varphi_i^k(\mathbf{r},t)\right|^2 \to \delta\left[\mathbf{r} - \mathbf{r}_i^k(t)\right], \qquad (13)$$

which yields from Eq.(12):

$$d_i(t) \propto \sum_{k=1}^{M} \mathbf{r}_i^k(t) . \tag{14}$$

Although Eq.(12) and Eq.(14) give close results for large number of walkers, the wavefunctions used in Eq.(12) offer much richer statistics and hence the predictions from Eq.(12) are expected to be more accurate that those of Eq.(14). This is so because the trajectories $\mathbf{r}_i^k(t)$ which participle in Eq.(14) are just separate sample points from the distributions $\left|\varphi_i^k(\mathbf{r},t)\right|^2$ of Eq.(12).

In order to test our method for electron dynamics in external fields we first compare in Fig.4 the results for the induced time dependent dipole moment (a), and the ionization (b), of 1D para-helium atom irradiated by a laser pulse with duration 3 fs at wavelength 300 nm (see the inset), and peak intensity 2.18 $10^{15}$ W/cm$^2$. We depict the results obtained from the TDQMC method (blue lines) together with the exact results, for $a=1$ a.u. and $b=1.5$ a.u. in Eq. (11) and Eq. (12). It is seen from Fig.4 that the TDQMC predictions for both the dipole moment and the ionization are very close to the exact results despite the high percentage of ionization which implies very strong deformations in the spatial distributions of the electron clouds. More important, however, is the comparison between the two predictions for ortho-helium where the outer electron is weakly bound and, in addition, there is an exchange interaction between the two electrons. Figure 5(a) shows the results for the same time parameters of the laser pulse as in Fig.4(a), but for much lower peak intensity of 5.6 $10^{13}$ W/cm$^2$. The blue and the red lines in Fig.5 show that using hole size $r_s=0.67$ a.u. in Eq.(12) ensures good correspondence with the exact dipole moment and ionization, while the green lines describe the case of zero hole size ($r_s=0$), which shows a very different time response. These results indicate that the accurate calculation of time evolution of a quantum system with correlated equal-spin electrons necessitates adequate description of the dynamic exchange effects, which can be comprehended within the proposed model.

In order to check the scaling properties of the TDQMC method we conducted initial tests of the parallel version of the TDQMC algorithm for the above examples on a massively parallel Blue Gene /P supercomputer. Since the different Monte Carlo walkers are guided by separate waves, the set of TDSE (Eq.5) and the guiding equations (Eq.(3)) are split and calculated in groups on different treads where the only communication between the processes occurs for calculating the effective Coulomb potential of Eq.(6). A scaling which is lower than quadratic with the number of walkers was found for one spatial dimension when increasing the number of processors from 128 up to 512 while keeping the same number of walkers for processor. It is important to stress that for large number of walkers the Monte Carlo sums of Eq.(6) and Eq.(7) can be restricted to a limited but sufficient number of walkers which may be, however, substantially smaller than the total number of walkers. The scaling for fixed number of walkers and increasing number of electrons is close to linear.

**Conclusions**

Here we present an approach to quantum dynamics where the evolution is described by ensembles of classical particles and quantum waves coupled through guiding equations. The guide waves obey a set of coupled Schrödinger equations where the use of effective potentials accounts for the local and nonlocal correlations between the electrons. We further improve the effective potentials to incorporate the effects of exchange interaction for fermionic states by explicitly introducing the size of the exchange hole as a screening parameter in the Coulomb interaction between the electrons. Our calculations for one-dimensional para- and ortho-helium atoms reveal that the exchange effects are correctly taken into account in our model for both the ground state and in a strong ionization regime where significant deformations of the electron cloud occur. The electron-pair density function calculated for the ground state electrons correctly describe the effects of exchange and Coulombic correlations. Although all quantities of interest can be expressed entirely in terms of walker configurations (without integrals calculation) the accuracy of the predictions can be improved further using the total distributions provided by the modulus square of the guide waves.

Finally, we should note that the quantum Monte Carlo methods have received some criticism in the literature in that the computing time for the *numerically exact* solution has to increase exponentially with the number of particles for fermions, which reflects the "sign problem" [27]. We have evidenced here that the TDQMC method can offer a controlled-accuracy *approximate* solution which scales as a low-order polynomial in time for both bosons and fermions. This scaling behavior occurs owing to the strategy employed: first split the quantum many-body problem to coupled single-body problems; second, use sampling particles (walkers) with well defined trajectories to couple the single-body quantum equations, and third, use an efficient procedure to map the discrete-to-continuous distributions. This allows us to accurately calculate correlated quantum dynamics while using relatively small number of walkers (hundreds to few thousands), where these walkers sample the much richer distributions provided by modulus square of the corresponding guide waves. Therefore, all the information about the nodal regions in case of fermions is carried by the evolving guide waves, which as a result, alleviates the necessity for using huge number of walkers. Further reduction in the computational workload can be expected using the Lagrangian 'go with the flow' viewpoint [28].

## Acknowledgments

The author gratefully acknowledges support from the National Science Fund of Bulgaria under contract DO-02-115/2008 "Centre of Excellence on Supercomputing Applications ( SuperCA)".

**Figure captions:**

**Figure 1.** Walker distribution in configuration space for para-helium (a) and for ortho-helium (b). The axes are in atomic units.

**Figure 2.** Nonlocal correlation lengths $\sigma_1$ and $\sigma_2$ of the two electrons, for para-helium (a) and for ortho-helium (b).

**Figure 3.** Electron-pair density function for the correlated ground state (blue lines) and from Hartree-Fock approximation (red lines), for para-helium (a) and for ortho-helium (b).

**Figure 4.** Induced dipole moment (a) and ionization (b) for para-helium atom (bosonic state). Blue lines – TDQMC result; red lines – exact result. The inset shows the shape of the external electric field.

**Figure 5.** Induced dipole moment (a) and ionization (b) for ortho-helium atom (fermionic state). Blue lines – TDQMC result; red lines – exact resut; green lines – for zero hole size $r_s = 0$.

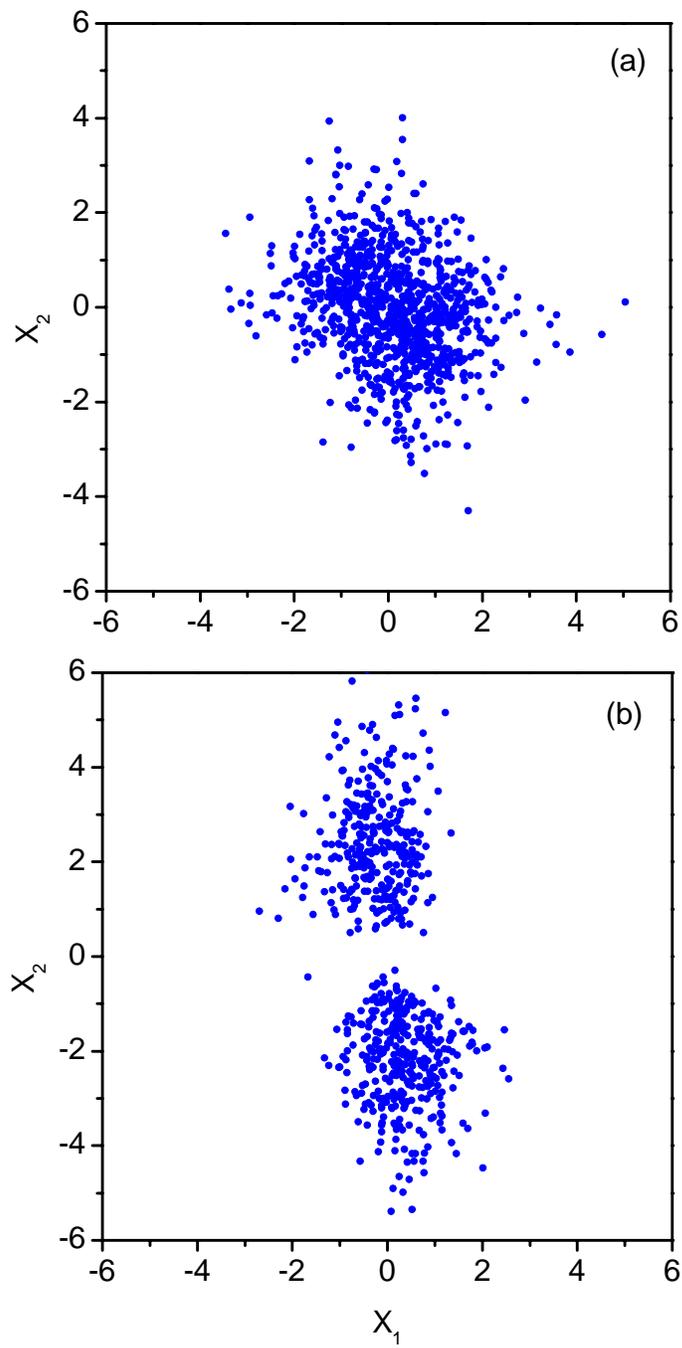

Figure 1

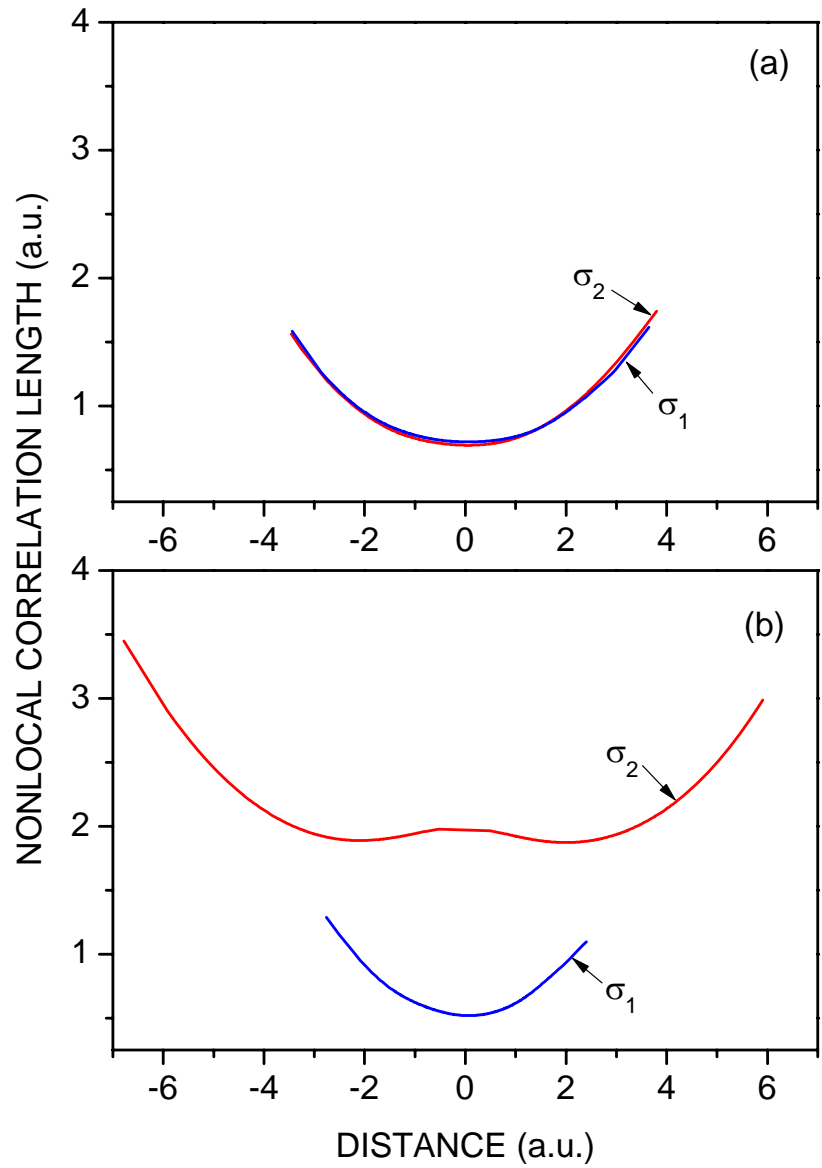

Figure 2

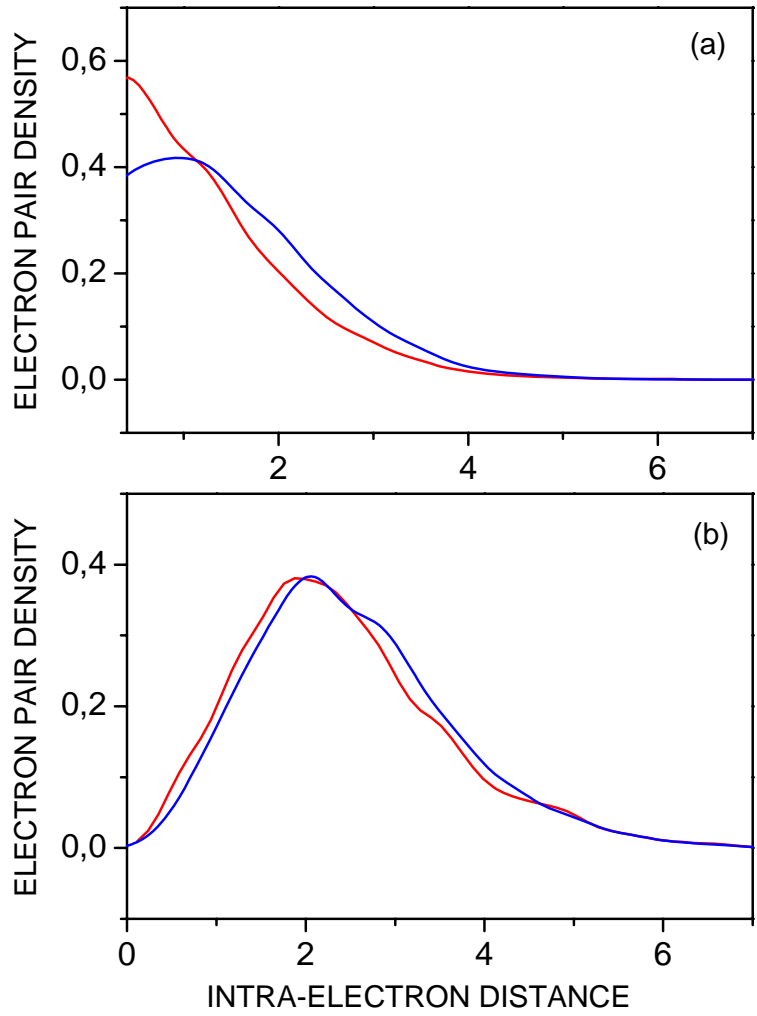

Figure 3

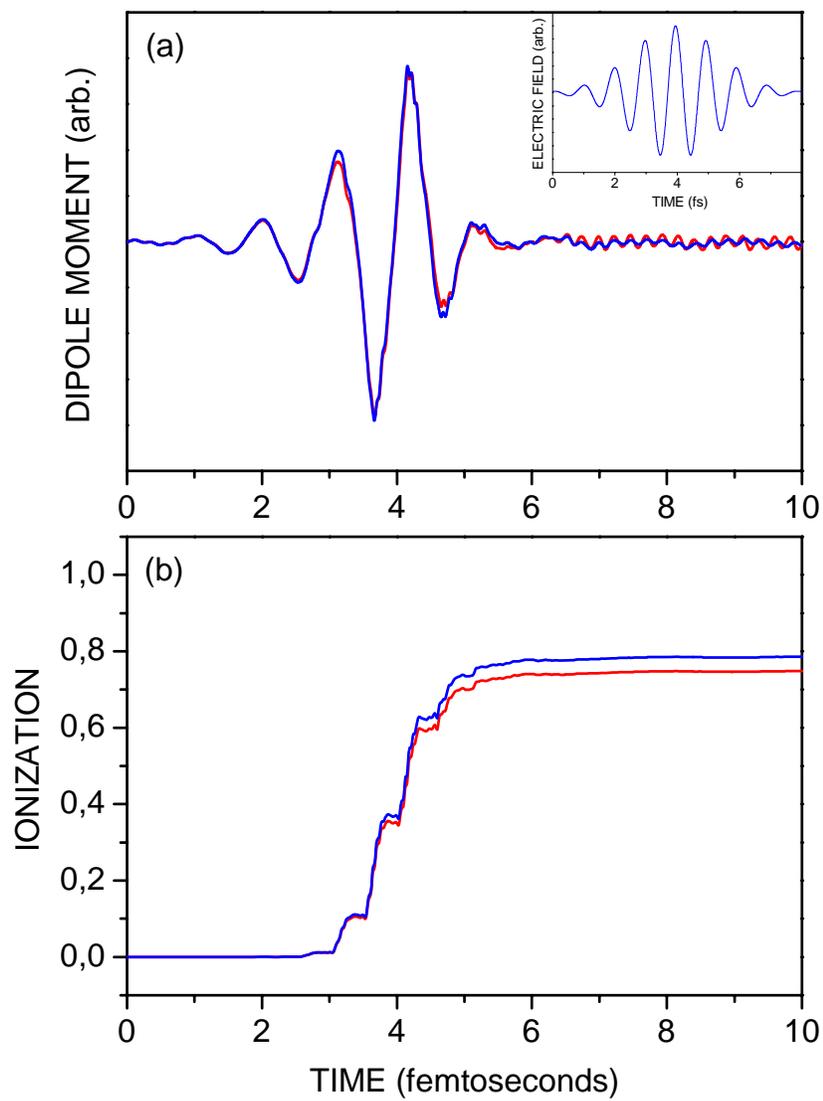

Figure 4

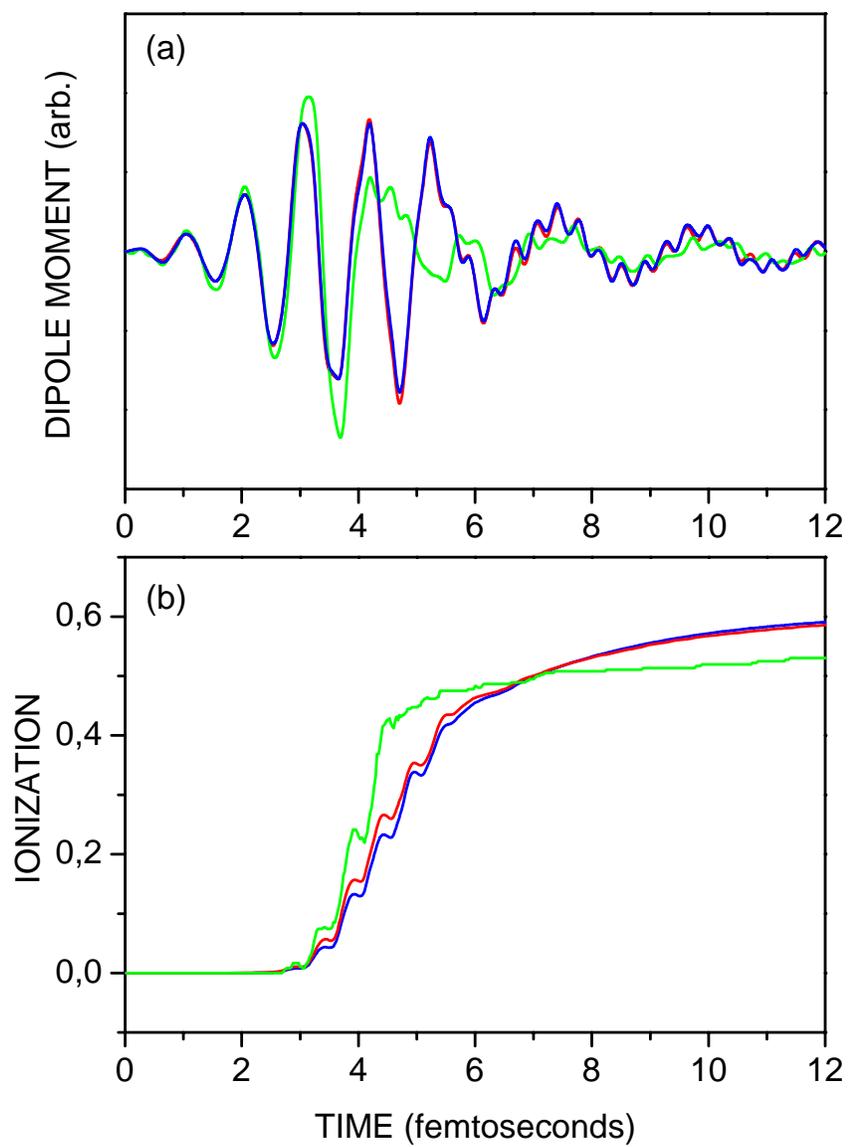

Figure 5